\documentclass{article}
\usepackage{graphicx} 
\usepackage[table]{xcolor}
\usepackage{color, colortbl}

\usepackage{authblk}
\providecommand{\keywords}[1]{\textbf{\textit{Keywords:}} #1}

\usepackage[ruled,vlined,linesnumbered]{algorithm2e} 
\usepackage{amsmath}   
\usepackage{amssymb}
\usepackage{mathtools}

\usepackage{graphicx}  
\usepackage{array}     
\usepackage{multirow}
\usepackage{pifont} 
\usepackage{makecell}

\newcommand{\xmark}{\textcolor{red}{\ding{55}}}  

\usepackage{booktabs}    
\usepackage{float} 

\usepackage{tikz}
\usetikzlibrary{shapes.geometric, arrows.meta, positioning}

\tikzstyle{block} = [rectangle, draw, fill=blue!10, text width=5.2em, align=center, minimum height=2.5em]
\tikzstyle{layer} = [rectangle, draw, fill=gray!10, rounded corners, text width=10em, align=center, minimum height=4.5em, minimum width=18em]
\tikzstyle{arrow} = [->, thick]

\title{Singularity Cipher: A Topology-Driven Cryptographic Scheme Based on Visual Paradox and Klein Bottle Illusions} 
\author[1]{Abraham Itzhak Weinberg}
\affil[1]{AI-WEINBERG, AI Experts, Tel Aviv, Israel, aviw2010@gmail.com}

\begin{document}
\maketitle
\begin{abstract}
This paper presents the Singularity Cipher, a novel cryptographic-steganographic framework that integrates topological transformations and visual paradoxes to achieve multidimensional security. Inspired by the non-orientable properties of the Klein bottle—constructed from two Möbius strips—the cipher applies symbolic twist functions to simulate topological traversal, producing high confusion and diffusion in the ciphertext. The resulting binary data is then encoded using perceptual illusions, such as the missing square paradox, to visually obscure the presence of encrypted content.
Unlike conventional ciphers that rely solely on algebraic complexity, the Singularity Cipher introduces a dual-layer approach: symbolic encryption rooted in topology and visual steganography designed for human cognitive ambiguity. This combination enhances both cryptographic strength and detection resistance, making it well-suited for secure communication, watermarking, and plausible deniability in adversarial environments. The paper formalizes the architecture, provides encryption and decryption algorithms, evaluates security properties, and compares the method against classical, post-quantum, and steganographic approaches. Potential applications and future research directions are also discussed.
\end{abstract}
\keywords{Topological Cryptography, Visual Steganography, Klein Bottle Cipher, Post-Quantum Security, Cognitive Encryption}

\section{Introduction}
\label{sec:introduction}

In the modern digital landscape, cryptographic systems must balance not only mathematical strength but also resistance to increasingly subtle forms of analysis, such as pattern recognition, statistical inference, and even human visual inspection \cite{kendhe2013survey}. Traditional cryptosystems, such as Advanced Encryption Standard (AES) \cite{dworkin2001advanced}, Rivest–Shamir–Adleman (RSA) \cite{imam2021systematic}, and post-quantum candidates like Kyber \cite{maram2023post}, focus predominantly on algebraic hardness assumptions. However, they often lack perceptual or structural obfuscation mechanisms that would prevent a message from being detected in the first place \cite{mandal2022digital}. \\
Recent efforts to expand the cryptographic design space have included the integration of chaotic dynamics and data motion \cite{zhang2023chaos}. One notable example is the Database in motion Chaos Encryption (DaChE) Algorithm~\cite{weinberg2025dynamic}, which applies chaos theory to NoSQL database systems to introduce dynamic data transformation and enhance security through unpredictable structural evolution. This highlights a broader shift toward multidimensional and structural cryptographic models \cite{lv2020study}.\\
This paper introduces the Singularity Cipher, a novel hybrid cryptographic-steganographic approach that combines two rarely integrated domains: topological transformation and visual paradox encoding. Inspired by the geometric structure of the Klein bottle---a non-orientable surface \footnote{On non-orientable surfaces like the Möbius strip, the lack of a well-defined 'inside' and 'outside' makes it impossible to assign a consistent directional orientation to the boundary.} formed by gluing together two Möbius strips \cite{putz2020topological}---the cipher employs symbolic twist functions to simulate traversal over such a surface, thereby achieving nonlinear confusion and diffusion of plaintext characters. \\
Following this symbolic encryption, the resulting ciphertext is mapped into a binary form and visually encoded using geometrically paradoxical figures such as the ``missing square illusion'' \footnote{The missing square puzzle is an optical illusion that exploits the unreliability of visual perception in geometric analysis, making it an ideal candidate for encoding information that appears visually consistent but contains hidden contradictions.} \cite{ninio2014geometrical}. These illusions not only obscure the presence of data but also exploit cognitive biases in human perception \cite{pohl2022cognitive}, introducing a layer of plausible deniability and visual complexity that traditional steganography methods often lack \cite{anderson1998limits}.\\
By combining the mathematical richness of non-orientable topological spaces \cite{fuladi2024short} with the deceptive power of visual illusions \cite{jiao2021image}, the Singularity Cipher aims to extend the cryptographic landscape into new multidimensional territory. While the scheme is not a replacement for established post-quantum cryptography \cite{joseph2022transitioning}, it offers a unique augmentation: an encryption paradigm where visual structure, spatial logic, and topological transformation serve as active components in data protection and concealment. This aligns with emerging frameworks like QUASAR~\cite{weinberg2025preparing}, which advocate for adaptable, forward-looking architectures to manage quantum-era risks, emphasizing the need for diversified strategies beyond algebraic hardness.\\
This paper formalizes the construction of the Singularity Cipher, presents a detailed algorithm, discusses its potential integration with post-quantum cryptographic primitives, and compares it to existing cryptographic and steganographic techniques in terms of functionality, obscurity, and resistance to visual and analytical attacks \cite{shamir1998visual}.

\subsection{Definition: Singularity Cipher}

Singularity Cipher is a novel hybrid cryptographic-steganographic scheme that encodes and encrypts information through a dual-layer system inspired by:
\begin{enumerate}
    \item \textbf{Topological transformations}, particularly Möbius strip twists and Klein bottle traversal \cite{clack2019distributed};
    \item \textbf{Geometric-paradox-based visual encoding}, such as the missing square illusion \cite{ponticorvo2010encoding}.
\end{enumerate}
It combines a symbolic twist-based cipher with a visual illusion-based bit encoding to obscure both the presence and the semantic content of a message.

\subsubsection*{Structure}

The cipher operates in two interconnected layers:

\begin{description}
    \item[1. Topological Cipher Layer (Klein Bottle Simulation):] 
    Each symbol $c \in \mathcal{A}$ from a finite alphabet is processed through two Möbius-style modular transformation functions $T_1$ and $T_2$, such that:
    \[
        E(c) = T_2(T_1(c)), \quad D(c) = T_1^{-1}(T_2^{-1}(c))
    \]
    This models traversal on a Klein bottle, where orientation and position change non-trivially due to non-orientability \cite{putz2020topological}.
    
    \item[2. Paradox Encoding Layer (Visual Steganography):] 
    Binary representations of encrypted data are visually encoded using geometric illusions \cite{ponticorvo2010encoding}:
    \begin{itemize}
        \item A \texttt{0}-bit is represented by a standard geometric arrangement (e.g., a valid triangle).
        \item A \texttt{1}-bit is represented by a paradoxical arrangement (e.g., a missing square illusion).
    \end{itemize}
\end{description}

\subsubsection*{Features}
\begin{itemize}
    \item Nonlinear, reversible symbol transformation inspired by topological surfaces \cite{yao2018topological}.
    \item Visual obfuscation of data via perceptual paradoxes \cite{chen2021perceptual}.
    \item Supports human-verifiable decryption or puzzle-based interfaces \cite{yang2016design}.
    \item Can be layered with post-quantum cryptography for added obfuscation \cite{edwards2020review}.
\end{itemize}

\subsubsection*{Purpose}

Singularity Cipher is designed to:
\begin{itemize}
    \item Provide an alternative encryption model based on topological and perceptual complexity \cite{polovnikov2020core}.
    \item Serve as a covert communication method through visual or physical media \cite{makhdoom2022comprehensive}.
    \item Explore foundational ideas for future post-quantum steganographic methods \cite{gabriel2013post}.
\end{itemize}

\subsection{Uniqueness of This Work}

The Singularity Cipher offers a distinctly novel approach to data protection by fusing two rarely connected paradigms: symbolic topological transformations and visual cognitive illusions \cite{myers2023topfusion}. Unlike conventional encryption algorithms that operate purely on numeric or algebraic operations \cite{yegireddi2016survey}, this cipher simulates traversal over a Klein bottle—a non-orientable topological surface—using key-based twist functions. Furthermore, it encodes the resulting ciphertext into a visual domain through geometric paradoxes that deceive the observer's perception \cite{ryu2018perception}.\\
This dual-layer mechanism offers not only encryption strength but also stealth and ambiguity. While post-quantum schemes resist decryption by quantum computers \cite{kappler2022post}, they remain visible as encrypted content. In contrast, the Singularity Cipher hides in plain sight, challenging both machines and humans to detect that a message exists at all \cite{baluja2017hiding}. This combination of symbolic complexity, visual disguise, and cognitive misdirection defines a new dimension in secure communication and sets this work apart from prior cryptographic and steganographic efforts \cite{taha2019combination}.\\
To formally capture this multidimensional encryption framework, we coin the term Singularity Cipher to describe the integration of topological symbol manipulation with visual-paradox-based steganography. A formal definition of the Singularity Cipher is presented above.

\subsection*{Paper Organization}

The remainder of this paper is organized as follows. Section~\ref{sec:background} provides an overview of related work in cryptography, steganography, and topological methods. Section~\ref{sec:motivation} introduces the motivation and theoretical justification for the proposed approach. In Section~\ref{sec:architecture}, we describe the system architecture and methodology of the Singularity Cipher. Section~\ref{sec:algorithm} details the encryption and decryption algorithms. Section~\ref{sec:security} analyzes the security properties of the scheme, while Section~\ref{sec:comparison} compares it with existing cryptographic and steganographic approaches. Section~\ref{sec:applications} outlines practical applications and use cases. Finally, Section~\ref{sec:conclusion} concludes the paper and discusses the directions for future work.

\section{Background and Related Work}
\label{sec:background}

This section outlines prior work and theoretical foundations related to the Singularity Cipher, specifically in the domains of cryptography, steganography, topological data transformation, and visual paradoxes.

\subsection{Cryptography and Nonlinear Transformations}

Traditional symmetric-key ciphers such as AES rely heavily on substitution-permutation networks to achieve confusion and diffusion \cite{dhall2024cryptanalysis,qayyum2020chaos}. Post-quantum algorithms, including lattice-based cryptography (e.g., Kyber \cite{maram2023post}), code-based schemes \cite{overbeck2009code}, and multivariate polynomial systems \cite{dey2023progress}, focus on algebraic structures that are resistant to quantum algorithms like Shor's \cite{ugwuishiwu2020overview} or Grover's \cite{grassl2016applying}.\\
While effective against brute-force and quantum attacks, these systems do not typically address data obfuscation at the structural or perceptual level \cite{horvath2020cryptographic}. Research into chaotic maps and non-linear geometries for cryptographic use has been limited but growing \cite{zhang2023chaos}, with some work exploring toroidal and hyperbolic data spaces \cite{singh2024systematic,genccouglu2017cryptanalysis}.

\subsection{Steganography and Visual Encoding}

Steganography focuses on concealing the existence of communication rather than encrypting the message itself \cite{ke2018steganography}. Techniques range from Least Significant Bit (LSB) image embedding \cite{bansal2020survey} to more advanced transformations in the frequency or wavelet domains \cite{yadahalli2020implementation} \cite{joshi2012image}. However, many of these methods can be detected through statistical steganalysis \cite{de2024comprehensive,fridrich2002practical}.\\
The Singularity Cipher differs by using visual paradoxes—specifically illusions involving area and geometry—as a medium of binary encoding \cite{jiao2021image}. This form of visual encoding resists both automated detection and human suspicion by presenting data as part of a believable, familiar visual structure \cite{bresciani2015pitfalls}.

\subsection{Topological Structures in Computation}

Topological concepts have been used in quantum computing (e.g., topological qubits) \cite{stern2013topological} and in error-correcting codes \cite{de2022quantum,yao2012experimental}, but they are underutilized in classical cryptography \cite{song2025can}. The Möbius strip and Klein bottle are well-known non-orientable surfaces that exhibit unique traversal properties \cite{polthier2003imaging}. When mapped to symbolic data transformations, they provide an opportunity to design reversible but disorienting encryption functions \cite{micciancio2005adaptive}.\\
Some prior research has investigated the use of braid groups and knot theory in cryptographic contexts \cite{dehornoy2004braid,sconza2024knot}, leveraging their non-commutative properties for key exchange \cite{ko2000new}. However, these approaches differ from the current work, which uses topology not for algebraic hardness but for structural encryption and symbol flow disruption \cite{barthe2019symbolic}.

\subsection{Cognitive Illusions and Perceptual Security}

Visual illusions, such as the ``missing square'' paradox \cite{ninio2014geometrical} or Penrose triangle \cite{draper1978penrose}, exploit the human visual system's assumptions about geometry, depth, and area \cite{marr2010vision}. While these illusions have been studied extensively in psychology \cite{jonsson2022deceptive, coren2020seeing}, their application in security is rare \cite{hayashi2008use}. A few steganographic systems have used optical illusions as distraction mechanisms \cite{jiao2021image}, but none have formally encoded binary data within paradoxical constructs \cite{goldwasser2019paradoxical}.\\
The Singularity Cipher bridges this gap by directly encoding binary digits using illusions whose geometry is locally plausible but globally inconsistent \cite{abbe2008local}. This creates a system where information is protected not only mathematically but cognitively \cite{patel2000cognitive}.

\section{Motivation}
\label{sec:motivation}
As digital communication becomes increasingly ubiquitous, so does the means and sophistication of adversarial analysis \cite{arcos2021digital}. Traditional cryptographic systems, while mathematically sound, are often optimized solely for computational hardness and do not address deeper structural or perceptual vulnerabilities \cite{vignesh2009limitations}. Adversaries may not always aim to break a cipher directly; instead, they may seek to detect the presence of encrypted communication, classify it, or infer its metadata through statistical, visual, or behavioral cues \cite{basyoni2020traffic}.\\
In parallel, the development of quantum computing presents a growing threat to widely deployed cryptographic primitives \cite{faruk2022review}. While post-quantum cryptography offers strong alternatives grounded in lattice problems, error-correcting codes, or multivariate polynomials \cite{bernstein2017post}, it still generally conforms to classical notions of message representation and transmission.
There is an emerging need for encryption methods that go beyond mathematical complexity alone—methods that incorporate cognitive misdirection, geometric transformation, and visual ambiguity to protect not just the contents but also the existence of sensitive information \cite{ke2018steganography}.\\
The Singularity Cipher addresses this need through a dual-layer design that draws from distinct disciplines. The topological layer introduces algebraic complexity using the properties of non-orientable surfaces such as the Möbius strip and Klein bottle, known for their non-trivial paths and symmetry-breaking properties that provide cryptographic confusion and diffusion \cite{yao2018topological}. The visual paradox layer exploits cognitive and perceptual vulnerabilities by hiding binary information inside geometric illusions, leveraging limitations in both human and machine interpretation \cite{ibrahim2021overview,andrade2019cognitive} to make ciphertext not only encrypted but also visually obfuscated.\\
This multidimensional approach aims to redefine the boundaries of encryption by integrating principles from topology, cognitive science, and optical illusion into cryptographic design \cite{al2014multidisciplinary}. By obscuring not only what is encrypted, but whether anything is encrypted at all, the Singularity Cipher serves as a foundation for systems that are inherently stealthy, obfuscated, and structurally unpredictable—qualities that may prove critical in the post-quantum era and in adversarial environments where data must survive not just decryption attempts, but also detection \cite{bloch2015channel}.

\section{Singularity Cipher Architecture}
\label{sec:architecture}

The Singularity Cipher is a hybrid cryptographic-steganographic system that integrates both topological and perceptual obfuscation mechanisms across two sequential transformation layers \cite{jassim2019hybrid}. This section describes the end-to-end data transformation process and outlines the mathematical and visual logic underlying each stage.

\subsection{System Overview}

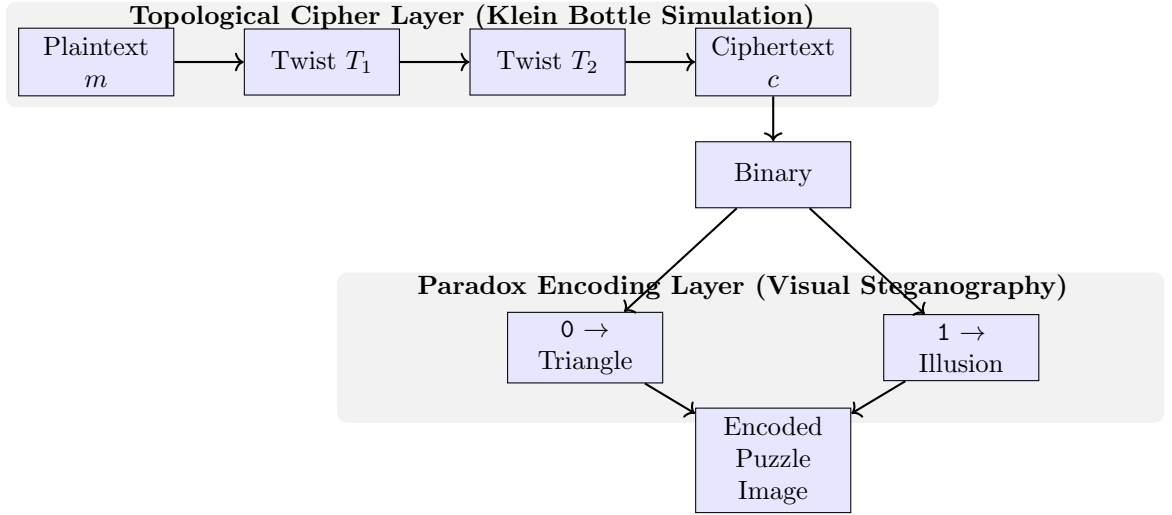
\begin{figure}[ht]
\centering
\begin{tikzpicture}[node distance=1.5cm and 2cm]
\fill[gray!10, rounded corners] (-6.2,3.2) rectangle (6.2,4.6);
\fill[gray!10, rounded corners] (-1.8,-1.0) rectangle (9.2,1.0);
\node at (0,4.4) {\textbf{Topological Cipher Layer (Klein Bottle Simulation)}};
\node at (3.6,0.8) {\textbf{Paradox Encoding Layer (Visual Steganography)}};
\node[block] (input) at (-5,3.8) {Plaintext $m$};
\node[block] (T1) at (-2,3.8) {Twist $T_1$};
\node[block] (T2) at (1,3.8) {Twist $T_2$};
\node[block] (enc) at (4,3.8) {Ciphertext $c$};
\draw[arrow] (input) -- (T1);
\draw[arrow] (T1) -- (T2);
\draw[arrow] (T2) -- (enc);
\node[block] (bin) at (4,2.3) {Binary};
\draw[arrow] (enc) -- (bin);
\node[block] (v0) at (1.5,0) {\texttt{0} $\rightarrow$ Triangle};
\node[block] (v1) at (6.5,0) {\texttt{1} $\rightarrow$ Illusion};
\draw[arrow] (bin) -- (v0);
\draw[arrow] (bin) -- (v1);
\node[block] (vis) at (4,-1.5) {Encoded Puzzle Image};
\draw[arrow] (v0) -- (vis);
\draw[arrow] (v1) -- (vis);
\end{tikzpicture}
\caption{Structure of the Singularity Cipher: a two-layer hybrid encryption scheme combining topological Möbius transformations and visual paradox encoding.}
\label{fig:singularity_cipher_diagram}
\end{figure}

Figure~\ref{fig:singularity_cipher_diagram} illustrates the overall structure of the proposed Singularity Cipher, which operates through two distinct layers:

\begin{enumerate}
    \item \textbf{Topological Cipher Layer}: This layer simulates a Klein bottle traversal by passing each plaintext symbol $m$ through two Möbius-style modular transformations, $T_1$ and $T_2$ \cite{putz2020topological,arnold2008mobius}. These transformations are key-dependent cyclic mappings over a symbol alphabet $\mathcal{A}$ that introduce orientation confusion and nonlinear diffusion \cite{qayyum2020chaos}. The resulting ciphertext $c$ is structurally resistant to simple inversion without knowledge of both twist parameters.
    
    \item \textbf{Paradox Encoding Layer}: The encrypted message $c$ is converted to its binary representation and mapped onto visual elements using perceptual encoding \cite{rauschenberger2006perceptual}. Each binary digit is rendered as a geometric object: a standard triangle for bit `0', or a paradoxical triangle (e.g., the "missing square" illusion) for bit `1' \cite{ninio2014geometrical,jonsson2022deceptive}. This enables steganographic embedding in diagrams, puzzles, or illustrations while leveraging human visual perception ambiguity as concealment \cite{xiang2019visual}.
\end{enumerate}

The final output is a composite visual image that not only hides the original message but also embeds structural ambiguity, making detection and decoding by adversaries more complex \cite{cruse2001crypto}. The dual-layer approach provides redundant security mechanisms—even if the visual encoding is detected, the underlying topological encryption maintains message protection \cite{lord2013designing}.

\subsection{Topological Cipher Layer: Klein Bottle Simulation}

The first layer introduces algebraic and spatial complexity through two Möbius-style permutation functions, $T_1$ and $T_2$, defined over a finite alphabet $\mathcal{A}$ \cite{li2008general,tesler2000matchings}:

\begin{equation}
    E(c) = T_2(T_1(m)), \quad D(c) = T_1^{-1}(T_2^{-1}(c))
\end{equation}

Each twist function $T_i$ represents a key-dependent cyclic permutation that may include bitwise reversal, modular shifting, and position-dependent character scrambling \cite{hehn2008permutation, ye2010image}. The non-orientable properties of the Klein bottle are simulated by ensuring that applying both $T_1$ and $T_2$ introduces irreversible confusion without access to both inverse functions \cite{polthier2003imaging}. The traversal mimics the Klein bottle's global self-intersection by causing symbol paths to cross or invert depending on position and context \cite{chas2012self}.

\subsection{Paradox Encoding Layer: Binary-to-Visual Mapping}

The ciphertext $c$ is converted to its binary representation $b$, and each bit is embedded in a visual structure according to the following encoding scheme \cite{lin2012based}:

\begin{itemize}
    \item A bit value of \texttt{0} is represented by a standard geometric shape, such as a consistent right triangle.
    \item A bit value of \texttt{1} is represented by a paradoxical image, such as the "missing square" illusion, where the area appears unchanged after rearrangement, despite the presence of a hidden void \cite{dabrowski2013framework}.
\end{itemize}

These representations are selected for their cognitive plausibility; the visual patterns appear structurally valid but contain implicit contradictions that conceal the data while maintaining plausible deniability \cite{halevi2005plausible,chapman2002plausible}.

\subsection{Visual Assembly and Output Generation}

Once the entire binary sequence has been transformed into visual symbols, the output is assembled into a single composite image or diagram \cite{dani1995automated}. The resulting image can be embedded in physical print, digital documents, or multimedia contexts, enabling both transmission and passive concealment \cite{shehab2022comprehensive}.\\
The system supports both automated and human-decodable decryption, where an observer with prior knowledge of the cipher rules can visually interpret and reconstruct the original binary stream from the image \cite{khoo2017human,jena2008survey}. The encoded message is thus protected by both mathematical permutation and perceptual masking, creating a cryptographic method that operates simultaneously in computational and perceptual domains \cite{mahalingam2023dual}.

\section{Algorithm Description}
\label{sec:algorithm}

This section formally defines the core procedures of the Singularity Cipher: the encryption and decryption algorithms. Each message is first passed through a symbolic topological transformation layer and then visually encoded using geometric paradoxes.\\
Let the plaintext be denoted as $m \in \mathcal{A}^*$, where $\mathcal{A}$ is a finite alphabet. Let $K = (k_1, k_2)$ represent the encryption key, consisting of two permutation parameters defining twist functions $T_1$ and $T_2$.

\subsection{Encryption Algorithm}

\begin{algorithm}[H]
\DontPrintSemicolon
\SetAlgoLined
\KwIn{Plaintext $m$, Key $K = (k_1, k_2)$}
\KwOut{Visual Cipher Image $\mathcal{I}$}
\BlankLine

Apply Möbius-style transformation:\;
$c_1 \leftarrow T_1(m, k_1)$ \tcp*{First twist (symbolic reordering)}
$c_2 \leftarrow T_2(c_1, k_2)$ \tcp*{Second twist (non-orientable traversal)}

Convert to binary:\;
$b \leftarrow \texttt{toBinary}(c_2)$

\ForEach{bit $b_i$ in $b$}{
    \eIf{$b_i = 0$}{
        $v_i \leftarrow \texttt{RenderTriangle}()$ \tcp*{Normal triangle}
    }{
        $v_i \leftarrow \texttt{RenderParadoxIllusion}()$ \tcp*{Geometric illusion (e.g., missing area)}
    }
}

$\mathcal{I} \leftarrow \texttt{AssembleImage}(v_1, v_2, \dots, v_n)$ \;
\Return{$\mathcal{I}$}
\caption{Singularity Cipher Encryption}
\label{alg:encrypt}
\end{algorithm}

\subsection{Decryption Algorithm}

\begin{algorithm}[H]
\DontPrintSemicolon
\SetAlgoLined
\KwIn{Visual Cipher Image $\mathcal{I}$, Key $K = (k_1, k_2)$}
\KwOut{Recovered Plaintext $m$}
\BlankLine

Extract binary string from visual symbols:\;
$b \leftarrow \texttt{DecodeBinaryFromImage}(\mathcal{I})$

Convert binary to ciphertext symbols:\;
$c_2 \leftarrow \texttt{fromBinary}(b)$

Apply inverse topological transformation:\;
$c_1 \leftarrow T_2^{-1}(c_2, k_2)$\;
$m \leftarrow T_1^{-1}(c_1, k_1)$

\Return{$m$}
\caption{Singularity Cipher Decryption}
\label{alg:decrypt}
\end{algorithm}

The decryption process reverses the encryption procedure by first analyzing the visual cipher image to extract embedded binary information, and then applying inverse topological transformations to recover the original plaintext. The algorithm begins by examining each visual element in the cipher image to distinguish between normal geometric shapes (representing binary 0s) and paradoxical illusions (representing binary 1s). This binary string is then converted back to symbolic form and subjected to inverse Möbius-style transformations, applied in reverse order to systematically undo the topological scrambling performed during encryption.

\section{Security Analysis}
\label{sec:security}
The security of the Singularity Cipher derives from a combination of symbolic transformation using topological permutation functions and perceptual encoding via geometric paradoxes \cite{yao2021new,vargas2016perceptual}. This section analyzes its resilience against classical and perceptual attacks.

\subsection{Confusion and Diffusion}
The topological cipher layer simulates traversal over a non-orientable surface (the Klein bottle), using two key-dependent twist functions $T_1$ and $T_2$ \cite{fuladi2024short}. These transformations ensure both confusion and diffusion \cite{qayyum2020chaos}:
\begin{itemize}
    \item \textbf{Confusion}: Symbol relationships are obscured due to position-dependent permutation and inversion patterns \cite{behnia2009cryptography}. The non-orientable nature of the Klein bottle makes it difficult to trace the original symbol flow \cite{stone2014metaphors}.
    \item \textbf{Diffusion}: A single-bit change in the input propagates through both twists, altering multiple output bits due to the nonlinear nature of the permutations \cite{liu2021image,wu2018cryptanalysis}.
\end{itemize}
These properties mirror those sought in modern block ciphers, such as avalanche effect and key sensitivity \cite{ramanujam2011designing,bogdanov2010analysis}.

\subsection{Key Space and Resistance to Brute Force}
The cipher uses two keys $(k_1, k_2)$, each defining a permutation of the message space. For an alphabet $\mathcal{A}$ of size $n$, the number of possible permutations is $n!$, making the combined key space size $(n!)^2$ \cite{bogdanov2012key}.

Even for small alphabets (e.g., $n = 256$), the key space becomes computationally infeasible to exhaust via brute-force search, ensuring high entropy and key unpredictability \cite{pospivsil2012lightweight,zhang2008cryptanalysis}.

\subsection{Resistance to Known Plaintext and Ciphertext-Only Attacks}
In a ciphertext-only attack, the visual representation of the ciphertext offers little statistical regularity due to the paradox encoding \cite{jiao2019known,gao2018methods}. Known plaintext attacks are also challenging because \cite{hariss2022towards}:
\begin{itemize}
    \item The symbolic transformation is nonlinear and reversible only with both $k_1$ and $k_2$ \cite{hussain2013literature}.
    \item The binary encoding obfuscates character boundaries \cite{popov2007binary}.
    \item The visual paradoxes disguise data presence, thwarting alignment-based inference \cite{bagdasaryan2024adversarial}.
\end{itemize}

\subsection{Steganographic Robustness}
Unlike LSB or frequency-domain steganography, the paradox encoding layer does not embed data in noise-prone image features \cite{bansal2020survey,joshi2012image}. Instead, it hides data in the semantics of the image itself, using illusions that appear innocuous but carry structured meaning to the receiver \cite{huo2024image,jiao2021image}.\\

This offers resistance to:
\begin{itemize}
    \item \textbf{Statistical steganalysis}, which typically relies on detecting minor image perturbations \cite{fridrich2002practical,agarwal2020image}.
    \item \textbf{Visual inspection}, where the illusion camouflages the presence of any information at all \cite{fridrich2002practical,troscianko2009camouflage}.
\end{itemize}

\subsection{Limitations and Assumptions}
\begin{itemize}
    \item The security of the cipher depends on the secrecy of the transformation keys and the correct rendering of paradox images \cite{dagdelen2013cryptographic,mabry2007unicode}.
    \item Optical distortions, compression artifacts, or automatic image filtering may degrade the ability to decode symbols reliably \cite{kim2021light,gschwandtner2007transmission,wang2004shield}.
    \item The visual encoding assumes that the receiver can correctly interpret geometric illusions; for machine interpretation, trained models may be needed \cite{westheimer2008illusions,hu2020towards}.
\end{itemize}

\section{Comparison with Existing Approaches}
\label{sec:comparison}

To contextualize the Singularity Cipher within the broader landscape of cryptographic and steganographic research, we compare its characteristics against several representative systems, including classical encryption schemes, post-quantum algorithms, and visual steganography methods \cite{sarveswaran2021cryptography,mandal2022digital}. This analysis reveals the unique positioning of our approach at the intersection of cryptographic security and perceptual concealment.

\subsection{Evaluation Framework}

We evaluate cryptographic and steganographic systems across five key dimensions that capture both traditional security properties and novel perceptual characteristics \cite{ramakrishna2024comprehensive}:

\begin{itemize}
    \item \textbf{Security Basis}: The underlying mathematical or perceptual foundation providing cryptographic strength \cite{lindell2017tutorials}.
    \item \textbf{Quantum Resistance}: Robustness against quantum attacks, particularly Shor's and Grover's algorithms \cite{ugwuishiwu2020overview,grassl2016applying}.
    \item \textbf{Visual/Perceptual Layer}: Integration of human perception mechanisms for additional security through cognitive concealment \cite{yang2023illusion}.
    \item \textbf{Steganographic Capability}: Ability to hide the existence of encrypted data within seemingly innocent visual content \cite{ke2018steganography}.
    \item \textbf{Confusion and Diffusion}: Effectiveness in obscuring input-output relationships through algebraic or topological transformations \cite{qayyum2020chaos}.
\end{itemize}

\subsection{Comparative Analysis}

Table \ref{tab:comprehensive_comparison} presents a comprehensive comparison of the Singularity Cipher against existing cryptographic and steganographic approaches, highlighting the distinctive features of each method.

\begin{table}[!ht]
\centering
\caption{Comprehensive Comparison of Cryptographic and Steganographic Approaches}
\label{tab:comprehensive_comparison}
\begin{tabular}{|p{3cm}|p{3.3cm}|p{2.0cm}|p{3.1cm}|p{2.6cm}|}
\hline
\textbf{Approach} & \textbf{Security Basis} & \textbf{Quantum Resistance} & \textbf{Visual/Perceptual Layer} & \textbf{Steganographic Use} \\
\hline
\textbf{Singularity Cipher (ours)} & Topological illusion, Möbius/Klein transformation & \makecell[c]{Unproven} & \multicolumn{1}{c|}{\textcolor{green}{\checkmark}} & \multicolumn{1}{c|}{\textcolor{green}{\checkmark}} \\
\hline
AES (Symmetric) & Block cipher, substitution-permutation & \multicolumn{1}{c|}{\xmark} & \multicolumn{1}{c|}{\xmark} & \multicolumn{1}{c|}{\xmark} \\
\hline
RSA / ECC & Integer factorization, elliptic curves & \multicolumn{1}{c|}{\xmark} & \multicolumn{1}{c|}{\xmark} & \multicolumn{1}{c|}{\xmark} \\
\hline
Kyber (PQC) & Lattice (MLWE) & \multicolumn{1}{c|}{\textcolor{green}{\checkmark}} & \multicolumn{1}{c|}{\xmark} & \multicolumn{1}{c|}{\xmark} \\
\hline
FrodoKEM (PQC) & Standard lattice (LWE) & \multicolumn{1}{c|}{\textcolor{green}{\checkmark}} & \multicolumn{1}{c|}{\xmark} & \multicolumn{1}{c|}{\xmark} \\
\hline
LSB Steganography & Bit-plane manipulation & \multicolumn{1}{c|}{\xmark} & \multicolumn{1}{c|}{\xmark} & \multicolumn{1}{c|}{\textcolor{green}{\checkmark}} \\
\hline
Visual Cryptography & Image-based XOR sharing & \multicolumn{1}{c|}{\xmark} & \multicolumn{1}{c|}{\textcolor{green}{\checkmark}} & \multicolumn{1}{c|}{\textcolor{green}{\checkmark}} \\
\hline
Braid Group Cryptography & Non-abelian algebra / topology & \multicolumn{1}{c|}{Partial} & \multicolumn{1}{c|}{\xmark} & \multicolumn{1}{c|}{\xmark} \\
\hline
Chaos-based Crypto & Dynamical systems / sensitivity & \multicolumn{1}{c|}{Partial} & \multicolumn{1}{c|}{\xmark} & \multicolumn{1}{c|}{\xmark} \\
\hline
Ambiguous Illusion Encoding & Human perception trick & \multicolumn{1}{c|}{\xmark} & \multicolumn{1}{c|}{\textcolor{green}{\checkmark}} & \multicolumn{1}{c|}{\textcolor{green}{\checkmark}} \\
\hline
\end{tabular}
\end{table}

\subsection{Discussion and Implications}

The comparative analysis reveals distinct advantages and limitations across different approaches. Classical encryption schemes like AES and RSA provide strong algebraic security through substitution-permutation networks and number-theoretic problems, respectively \cite{dworkin2001advanced,imam2021systematic}. However, they lack visual concealment capabilities and are vulnerable to quantum cryptanalysis \cite{mitra2017quantum}. Their high confusion and diffusion properties make them excellent for traditional cryptographic applications but render encrypted data easily identifiable \cite{abd2019classification}.\\
Post-quantum cryptographic algorithms such as Kyber and FrodoKEM address quantum vulnerabilities through lattice-based constructions, offering proven resistance against Shor's algorithm \cite{maram2023post,guo2020key,nejatollahi2019post}. Despite their quantum robustness, these methods operate within conventional algebraic frameworks and provide no mechanism for visual obfuscation or perceptual concealment \cite{hosseini2024comprehensive}.\\
Steganographic approaches like LSB manipulation excel at hiding data existence but offer minimal cryptographic protection \cite{bansal2020survey,dhawan2021analysis}. Visual cryptography provides both concealment and some level of security through image-based secret sharing, yet remains vulnerable to various image processing attacks and statistical analysis \cite{vyas2014review,talhaoui2025vulnerability}.\\
The Singularity Cipher occupies a unique position by integrating topological transformations with perceptual illusions, creating a dual-layer security model \cite{mahalingam2023dual}. This approach combines strong symbolic transformation through Möbius and Klein bottle mappings with cognitive concealment via paradox-based steganography \cite{peng1998second,yang2021linguistic}. While its quantum resistance remains theoretically unproven, the topological foundation suggests potential robustness against both classical and quantum attacks \cite{tzalenchuk2010towards}.\\
Furthermore, the Singularity Cipher's design philosophy enables hybrid deployment scenarios where it can be combined with established post-quantum algorithms \cite{fedorov2023deploying} \cite{scientific2025multilayered}. This layered approach would inherit the computational robustness of proven Post Quantum Cryptography (PQC) schemes while adding structural and visual stealth capabilities unique to our method \cite{mikic2025post}.\\
The analysis demonstrates that while the Singularity Cipher may not replace high-throughput traditional encryption systems, it offers compelling advantages for scenarios requiring covert, cognitively shielded communication \cite{makhdoom2022comprehensive,singh2024psychological}. Its integration of cryptographic strength with perceptual invisibility makes it particularly suitable for applications where both security and concealment are paramount, opening new avenues for secure communication in adversarial environments \cite{schaefer2015secure,shamsi2024visually}.

\section{Applications and Use Cases}
\label{sec:applications}
The Singularity Cipher offers a unique integration of symbolic encryption and visual steganography, making it particularly well-suited for environments that require both security and covert communication \cite{makhdoom2022comprehensive}. This section outlines key application domains where the cipher provides distinctive advantages.

\subsection{Covert Communication in High-Risk Environments}
In authoritarian regimes or conflict zones, the ability to hide not just the content but the very presence of encrypted communication is critical \cite{dal2022walking,chamales2011securing}. The Singularity Cipher allows messages to be embedded in seemingly innocent diagrams or illustrations, evading censorship and detection \cite{khan2007message}. Since the visual output resembles mathematical puzzles or artwork, it offers plausible deniability \cite{chapman2002plausible,johnson1998exploring}.

\subsection{Steganographic Watermarking for Intellectual Property Protection}
The visual paradox layer can be used to encode ownership or licensing metadata directly into scientific illustrations, technical drawings, or architectural blueprints \cite{evsutin2020digital,sengupta2019ip}. These hidden watermarks are difficult to detect and remove without specific knowledge of the encoding scheme, providing a novel form of content authentication \cite{kadian2021robust,muhammad2017image}.

\subsection{Secure Instruction Encoding in Printed or Visual Media}
In espionage or intelligence contexts, visual steganography allows instructions or data to be embedded in public media such as posters, packaging, or digital art \cite{zebari2020image,hall2014encoding}. The use of paradoxical geometry makes the encoded data accessible only to viewers with specific decoding instructions, adding a layer of obfuscation that is resistant to machine detection \cite{kamitani2005decoding,barkhi2024integrating}.

\subsection{Human-Perceptual Cryptography for Cognitive Interfaces}
The cipher could serve in secure human-machine interfaces where visual illusions trigger specific responses or actions \cite{lin2022touch,andrade2019cognitive}. For example, in Augmented Reality (AR) or heads-up displays, paradoxical encodings may serve as secure tokens that are recognized by trained human agents but remain undecipherable to automated systems \cite{syed2022depth,goyal2010founding,kaminska2018storing}.

\subsection{Augmenting Post-Quantum Encryption with Structural Obfuscation}
Though not a replacement for standard post-quantum cryptographic algorithms, the Singularity Cipher can augment them by encoding their outputs into paradoxical visual structures \cite{gordienko2025hnn,sengupta2020structural}. This provides additional defense in depth that combines computational intractability with perceptual stealth \cite{yendamury2021defense,mollicchi2017flatness}.

\section{Conclusion}
\label{sec:conclusion}

This paper introduced the Singularity Cipher, a novel hybrid encryption and steganographic scheme that integrates topological transformations inspired by the Klein bottle with visual paradox-based data encoding. By leveraging symbolic permutations across non-orientable geometries and embedding binary data in perceptually deceptive visual structures, the cipher offers multidimensional security that goes beyond traditional mathematical hardness assumptions.\\
We demonstrated that the cipher provides strong confusion and diffusion, a large key space, and enhanced stealth through cognitive and perceptual obfuscation. In contrast to classical and post-quantum cryptographic systems, which focus primarily on algorithmic complexity, the Singularity Cipher also addresses the growing need for concealment and plausible deniability in adversarial environments.\\
The architecture supports applications in covert communication, watermarking, and secure visual interfaces, particularly where data needs to remain hidden in plain sight. Moreover, the cipher can be combined with post-quantum primitives to create a multi-layered defense strategy that is both computationally secure and visually elusive.\\
Future work includes formalizing the cipher's resistance to machine learning-based steganalysis, developing automated visual encoders and decoders, and exploring other topological surfaces—such as higher-genus manifolds—as symbolic spaces for encryption. We also anticipate the potential for adaptation in augmented reality and human-centric cryptography, where visual cognition becomes part of the secure communication channel.


\bibliographystyle{IEEEtran}
\bibliography{ref.bib}

\end{document}